\documentclass[aps,prb,superscriptaddress, twocolumn, longbibliography]{revtex4-2}

\usepackage[english]{babel}
\usepackage{times}
\usepackage{graphicx}
\usepackage{graphics}
\usepackage{amsmath}
\usepackage{mathtools}
\usepackage{amsfonts}
\usepackage{amssymb}
\usepackage{dsfont}
\usepackage{epstopdf}
\usepackage{makeidx}
\usepackage{subfigure}
\usepackage{color}
\usepackage{pgf}
\usepackage{bm}
\usepackage{ulem}
\usepackage{tikz} 
\usepackage{hyperref}
\makeindex

\begin{document}

\title{Pairing amplification induced by nonadiabatic effects on the electron-phonon interaction throughout the BCS-BEC crossover}

\author{Victor Velasco} 

\affiliation{School of Pharmacy, Physics Unit, University of Camerino, Via Madonna delle Carceri 9, 62032 Camerino, Italy}

\author{Giovanni Midei}

\affiliation{School of Science and Technology, Physics Division, University of Camerino, Via Madonna delle Carceri, 9B, Camerino (MC), Italy}

\affiliation{INFN-Sezione di Perugia, 06123 Perugia, Italy}

\author{Massimo Capone}

\affiliation{International School for Advanced Studies (SISSA), Via Bonomea 265, I-34136 Trieste, Italy}
\affiliation{CNR-IOM Democritos, Via Bonomea 265, I-34136 Trieste, Italy}

\author{Andrea Perali} 

\affiliation{School of Pharmacy, Physics Unit, University of Camerino, Via Madonna delle Carceri 9, 62032 Camerino, Italy}

\begin{abstract}
Nonadiabatic effects in the electron-phonon coupling are important whenever the ratio between the phononic and the electronic energy scales, the adiabatic ratio, is non negligible. For superconducting systems, this gives rise to additional diagrams in the superconducting self-energy, the vertex and cross corrections. In this work we explore these corrections in a two-dimensional single-band system through the crossover between the weak-coupling BCS and strong-coupling Bose-Einstein regimes. By focusing on the pseudogap phase, we identify the parameter range in which the pairing amplitude is amplified by nonadiabatic effects and map them throughout the BCS-BEC crossover. These effects become stronger as the system is driven deeply in the crossover regime, for phonon frequencies of the order of the hopping energy and for large enough electron-phonon coupling. Finally, we provide the phase space regions in which the effects of nonadiabaticity are more relevant for unconventional superconductors.
\end{abstract}


\maketitle

\section{Introduction}

The Bardeen-Cooper-Schrieffer (BCS) theory describes the microscopic mechanism behind the superconducting phase in the weak electron-phonon coupling regime \cite{BCS1957}. In this context, the pioneering work of Migdal established that for conventional metals one is able to treat the diagrammatic series arising from the electron-phonon interaction in a perturbative manner. This is done via the expansion parameter $\lambda\omega_0/E_F$, where $\lambda$ is the electron-phonon coupling, $\omega_0$ is a typical phononic energy scale and $E_F$ is the Fermi energy. As a consequence of the Migdal's theorem, in the adiabatic regime, where $\omega_0/E_F \ll 1$, the series can be truncated in the first order of interaction  \cite{Migdal1958}, neglecting vertex corrections to the electron-phonon coupling. By extending this formalism to the strong coupling regime of the superconducting state, where $\lambda \approx 1$, one ends with the Migdal-Eliashberg (ME) equations that accurately describes conventional superconductors \cite{Eliashberg1960}.

While Migdal's theorem applies to low-temperature superconductors, its validity  can be questioned in more complex superconducting materials, including those showing high critical temperatures. In several of these compounds, due to strong correlations and/or band structure effects, the bands at the Fermi level are shallow, leading to a small $E_F$. Besides, phonon modes with larger $\omega_0$ can be involved in the pairing, or more generally in electron-phonon effects, and bosons of electronic origin can be relevant. 
An interesting example is provided by alkali-doped fullerene \cite{Hebard1991,Gunnarsson1997,Ganin2008,Capone2009}, in which narrow bands arising from molecular orbitals and relatively high-energy phonon modes conjure to give anomalously large values of the adiabatic ratio.
It has indeed been shown that the phase diagram of these compounds can be reproduced using an nonadiabatic approximation for the electron-phonon coupling \cite{Capone2002,Capone2009}, which compares well with a fully {\it ab-initio} theory including dynamical phonons \cite{Nomura2015,Nomura2016}.

Despite the universally accepted central role of electron-electron interactions, electron-phonon coupling shows anomalous signatures also in underdoped cuprates, where a large isotope effect, together with a mass dependent penetration length \cite{Uemura1991, Zhao1997, Damascelli2003, Lanzara2001}, points to a scenario where Migdal's theorem breaks down. Other remarkable examples of superconducting systems that are expected to display nonadiabatic effects include monolayer $\mathrm{FeSe}$ on $\mathrm{SrTiO_3}$ \cite{Wang2012, Lee2014, Liu2015}, sulfur hydrides \cite{Errea2015, Drozdov2015}, magic-angle-twisted bilayer graphene \cite{Cao2018} and doped $\mathrm{SrTiO_3}$ \cite{Marel2011}. This shows the importance of a deep understanding of vertex corrections to the electron-phonon coupling and their nonadiabatic effects.

The consequences of the breakdown of Migdal's theorem in the theory of the normal and superconducting state have been thoroughly analyzed in the context of the nonadiabatic theory of superconductivity. \cite{Grimaldi1995, Pietronero1995_1, Pietronero1995_2, Botti2002, Grimaldi99}. Moreover, these vertex corrections were also studied under the effects of band-filling \cite{Cappelluti2003, Perali1998, Cappelluti1996}, electronic correlation \cite{Boeri2003, Huang2003, Capp2004,Paci2005,Paci2006}, polaron formation \cite{Capone1998,Capone2003,Capone2006}, in Dirac materials \cite{Roy2014} and within dynamical mean-field theory \cite{Sangiovanni2006},
exploring their complex dependence on model parameters.  In this context, the generalization of the ME equations to include the first vertex corrections was developed \cite{Pietronero1995_2} and recently it was evaluated without further approximations \cite{Schrodi2020}.
On top of that, it was also claimed that unconventional pairing could 
be mediated solely by an isotropic electron-phonon coupling by solving 
the generalized ME equations \cite{Schrodi2021}, but this is in contradiction with quantum Monte Carlo calculations \cite{Dee2023}. Overall, it is clear that the physics introduced by these nonadiabatic
effects is rich and further exploration is necessary.


Irrespective of the dimensionality of the system and of the normal state, the BCS superconductivity of extended Cooper pairs and the Bose-Einstein regime of preformed molecular-like pairs are smoothly connected through the progressive reduction of the size of the pairs, when the inter-particle attraction (and/or the density \cite{Andrenacci1999}) is tuned. 
For this reason, in any dimension, the continuous evolution of a system of electrons as the attraction changes from weak to strong coupling has been commonly indicated in the literature as the BCS-BEC crossover, where BEC refers both to a condensation or a quasi-condensation of bosonic pairs.
This phenomenon has attracted great attention in different fields such as particle physics \cite{Baym1983,Yamamoto2007}, nuclear physics \cite {Baldo1995, Strinati2018} and condensed matter \cite{Micnas1990,Toschi2005a,Toschi2005b}. In the last decades, the BCS-BEC crossover has been widely studied in the context of ultra-cold quantum gases (see Refs. \cite{Inguscio2007} and \cite{Zwerger2012} for an experimental and theoretical overview, respectively) and, more recently, it has been observed varying the density in the single-band superconductor $\mathrm{Li_xZrNCl}$ \cite{Nakagawa2021}.
 Moreover, evidence of this phenomenon can be found in multi-gap and multi-band superconductors, where interesting interference effects can arise due to the coherent mixture of partial condensates forming in each band \cite{Eagles1969,Chen2005,Tajima2019, Tajima2020, Midei2023}. Evidences of the BCS–BEC crossover and strong coupling superconductivity have been reported in doped iron-chalcogenide superconductors \cite{Sato2018, Rinott2017, Kashara2014, Lubashevsky2012} and underdoped superconducting cuprates \cite{Velasco2021,Perali2000, Perali2002}. Additionally, $\mathrm{H_3 S}$ hydride superconductors near a Lifshitz transition are in regimes where Migdal's approximation is no longer valid and the coexistence of a BCS-BEC and a BCS condensates was proposed \cite{Bianconi2016}. 
 This scenario leaves little doubt about the interest in the understanding of the effects of nonadiabatic diagrams in the BCS-BEC crossover.

In this work, we analyze the nonadiabatic effects on the electron-phonon coupling throughout the BCS-BEC crossover. We first introduce the relevant Feynman diagrams including nonadiabatic corrections, {\it i.e.}, the vertex and cross diagrams in the superconducting self-energy. We report the analysis of their dependencies on exchanged momenta, chemical potential and phonon frequency by using a single-band two-dimensional tight-biding dispersion. The diagrams show a rich structure in the relevant parameter space, thus in order to understand their relevance to the BCS-BEC crossover, we solve the self-consistent gap and density equations at finite temperature within a mean-field BCS scheme. We can define a mean-field critical temperature, $T_{MF}$, which plays the role of a pseudogap temperature and a real renormalized superconducting critical temperature, which in two-dimensions consists of the Berezinskii–Kosterlitz–Thouless (BKT)  transition temperature, $T_{BKT}$. Our simplified approach amounts to work in the pseudogap phase by choosing a temperature $T$ such that $T_{BKT} < T < T_{MF}$. The normal state pseudogap can be described as a superconducting gap without coherence, transporting to finite temperature the possibility to tune the BCS-BEC crossover with the electron-phonon coupling, as demonstrated for ultracold fermion atoms \cite{Marsiglio2015, Perali2011, Gaebler2010}. 
In this way, we calculate the condensate fraction of incoherent pairs for finite temperatures, $\alpha$, and determine the boundaries of the BCS, crossover and BEC states of the system \cite{Guidini2014}, mapping the relevant parameters and calculating the associated correction to the effective interaction due to nonadiabatic effects. Remarkably, we show that these corrections can enhance the pairing throughout the whole crossover, with special attention to the crossover towards the BEC side, where the chemical potential is strongly renormalized and more than half of the particles are forming the condensate. Finally, we connect our numerical results with the parameters characterizing specific materials. The goal is to get new insights into the BCS-BEC crossover brought by the nonadiabatic corrections and to locate the range of parameters where these corrections are most relevant.

The paper is organized as follows. In Sec. \ref{Section: Models} we review the nonadiabatic theory of superconductivity and introduce the vertex and cross corrections to the electron-phonon coupling, which are responsible for the correction to the effective pairing of electrons. Then, we introduce the mean-field equations used to determine the boundaries of the BCS, crossover and BEC phases at finite temperature, where we identify the relevant regime of parameters for the study of the nonadiabatic effects. Section \ref{Section: numerical analysis} is devoted to the numerical analysis of the behavior of the vertex corrections and their relevance to the BCS-BEC crossover. Finally, in Sec. \ref{Section: Conclusion} we make our final remarks.

\section{Models}
\label{Section: Models}

\subsection{Beyond Migdal's theorem}

\begin{figure}
    \centering\includegraphics[width = \linewidth]{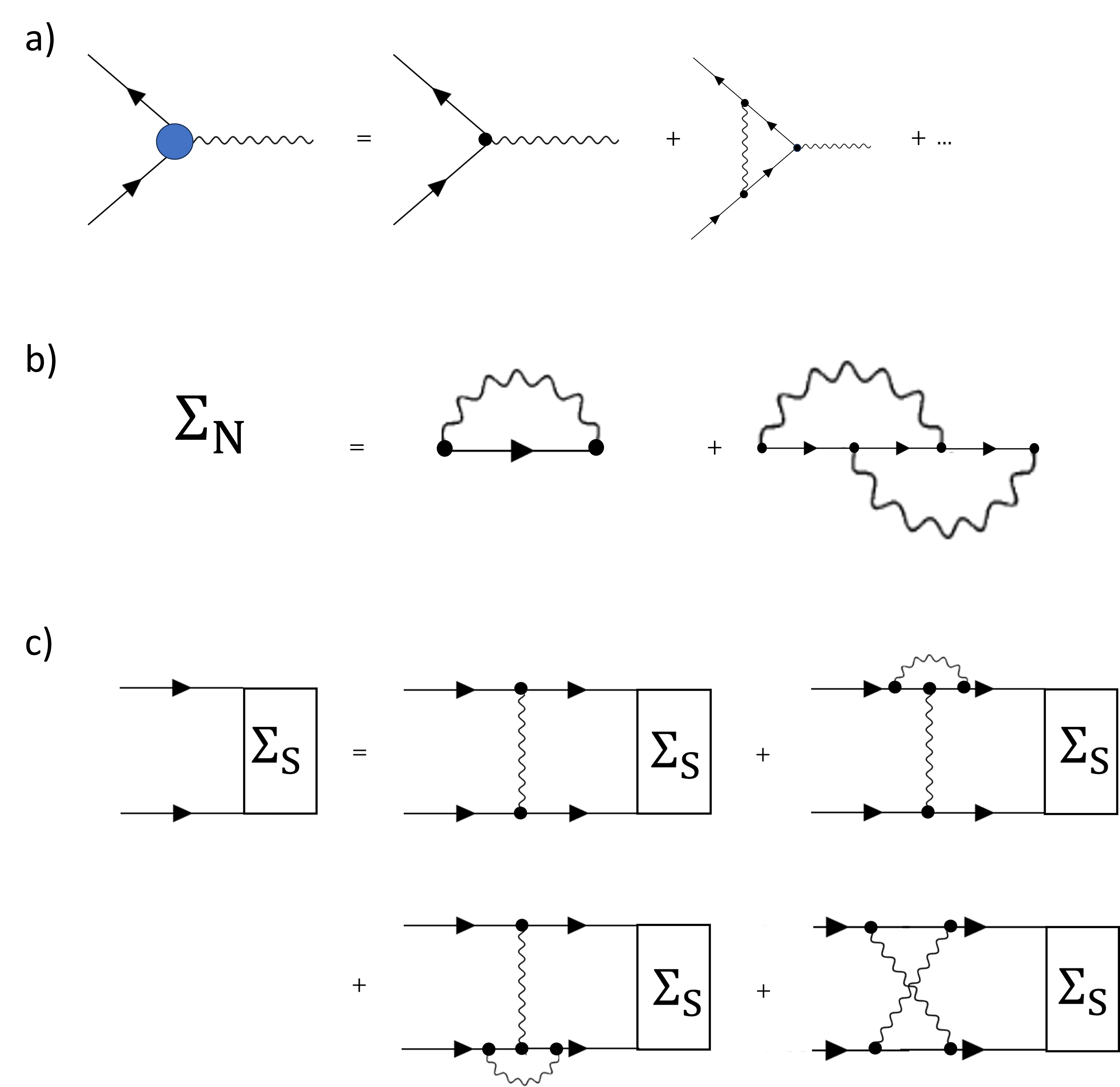}
    \caption{Diagrammatic representation of the vertex corrections. Electron and phonon propagators are  solid and wiggly lines, respectively, and the electron-phonon interaction, $g(\mathbf{q})$, is represented by a solid black circle. a) Electron-phonon interaction diagrams up to the first order in the corrections. The full vertex function, $\Gamma(\mathbf{q}, \omega)$, is represented by the blue dot. b) The normal state self-energy  corrected by the vertex diagram. c) The superconducting state self-energy. The nonadiabatic effects introduce the vertex (the second and third diagrams on the r.h.s.) and cross (last diagram) corrections to the self-energy.}
    \label{fig:diagrams}
\end{figure}

The nonadiabatic theory of superconductivity was thoroughly analyzed in different contexts and different mathematical derivations can be found in the literature. We introduce here the basic concepts that are relevant for the analysis of the BCS-BEC crossover. 
In Fig. \ref{fig:diagrams}a), we show the diagrammatic representation of the structure of the electron-phonon coupling up to the first order. All the difficulty in the evaluation of the diagrams descends from the structure of the vertex function, $\Gamma (\mathbf{q}, \omega)$, where $\mathbf{q}$ and $\omega$ are the exchanged phonon momenta and frequency, respectively. In his seminal work \cite{Migdal1958}, Migdal showed that all the non-zero corrections are proportional to $\lambda \omega_0/E_F$, setting the regime of validity of the Migdal's approximation to conventional superconductors, where $\omega_0/E_F$ is of the order of $10^{-3} - 10^{-4}$. However, in unconventional superconductors such as cuprates, fullerenes, $\mathrm{MgB_2}$ and others, this ratio is no longer negligible and the corrections must be taken into account. In this case, as shown in Fig. \ref{fig:diagrams}b), the normal-state electronic self-energy is corrected by the nonadiabatic contribution coming from the vertex correction and the self-energy for the superconducting state, as shown in Fig. \ref{fig:diagrams}c), gets also the contribution from the cross correction diagram \cite{Pietronero1995_1, Grimaldi1995, Pietronero1995_2}.

From the diagrams in Figs. \ref{fig:diagrams}b) and c), we get the equations for the diagonal, $\Sigma_N$, and off-diagonal, $\Sigma_S$, self-energies for the normal and superconducting states, respectively, which read 

\begin{eqnarray}
    \Sigma_N(k) & = & \sum_{k^\prime} V_0(k-k^\prime)\left[ 1 + P(k, k^\prime)\right]G(k^\prime), \label{Eq: self-energy_normal}\\
    \Sigma_S(k) & = & \sum_{k^\prime}V(k,k^\prime)G(k^\prime)G(-k^\prime)\Sigma_S(k^\prime).
    \label{Eq: self-energy_suc}
\end{eqnarray}
Here $k = (i\omega_k, \mathbf{k})$ is a short-hand notation for the summation over Matsubara frequencies and momenta. The bare electron-phonon interaction $V_0(k-k^\prime) = g^2 (\mathbf{k -k ^\prime})D_0 (k-k^\prime)$ is written in terms of the matrix elements for the electron-phonon coupling $g(\mathbf{k - k^\prime})$ and
\begin{equation}
G(k)= \frac{1}{i\omega_k - \xi(\mathbf{k}) - \Sigma_N(k)},
\end{equation}
which is the electronic Green's function written in terms of the electronic dispersion $\xi(\mathbf{k})$, and
\begin{equation}
    D_0(q) = \frac{-2\omega_0}{\omega_q^2 + \omega_0^2},
\end{equation}
is the bare phonon propagator, written in terms of a constant phonon frequency $\omega_0$ and the exchanged frequency $\omega_q$. In Eqs. \eqref{Eq: self-energy_normal} and \eqref{Eq: self-energy_suc} we define the vertex function

\begin{equation}
    P(k,k^\prime) = \sum_p V_0 (k - p)G(p - k + k^\prime)G(p), 
    \label{Eq: P}
\end{equation}
which renormalizes the interaction kernel appearing in $\Sigma_N(k)$.
The interaction kernel of the linearized superconducting gap equation is dressed by

\begin{equation}
    V(k,k ^\prime) = V_0(k-k^\prime)\left[ 1 + 2P(k,k^\prime)\right] + C(k,k^\prime),
    \label{Eq: effective}
\end{equation}
where $C(k,k^\prime)$ is the cross correction, which appears as the last diagram in the superconducting self-energy (see Fig. \ref{fig:diagrams}c)), and is written as

\begin{equation}
    C(k,k^\prime) = \sum_p V_0(k-p)V_0(p-k^\prime)G(p)G(p-k-k^\prime).
    \label{Eq: C}
\end{equation}
In the Migdal's limit, both $P(k,k^\prime)$ and $C(k,k^\prime)$ can be neglected and one recovers the usual self-energies and the ordinary Migdal-Eliashberg equations. We notice that, due to the construction of the kernel in Eq. \eqref{Eq: effective}, positive $P$ and negative $C$ amplify the pairing interaction. Both corrections have their own complicated structure in momenta and frequency space \cite{Pietronero1995_1, Pietronero1995_2}, mainly $P$ is dependent on the exchanged momenta $\mathbf{q = k - k^\prime}$ and $C$ on the sum $\mathbf{Q = k + k^\prime}$. In order to simplify the analysis, we neglect the self-energy effects in the electronic Green's functions appearing in $P(k,k^\prime)$ and $C(k,k^\prime)$ and use the bare normal state propagator $G_0(k)^{-1} = (i\omega_k - \xi(\mathbf{k}))$, in such a way that the Matsubara summations can be done analytically, giving a set of closed equations for the corrections \cite{Perali1998} and we are left only with the momentum summations, which we perform numerically.


\subsection{Finite temperature model}

In order to establish the regime of parameters that are relevant for the BCS-BEC crossover and thus study the nonadiabatic effects, we use information acquired from the study of the finite temperature coupled mean-field BCS equations for the superconducting gap and density for a single-band system. As we mentioned in the introduction, in this work we identify a pseudogap regime as a temperature range between the mean-field pairing temperature and the actual superconducting critical temperature of the two dimensional system, namely the BKT transition temperature (see below). We start with the finite-temperature gap equation, which is given by

\begin{equation}
\Delta(\mathbf{k})=-\frac{1}{\Omega} \sum_{\mathbf{k}^{\prime}} \frac{V\left(\mathbf{k}, \mathbf{k}^{\prime}\right) \tanh\left({\frac{\beta E(\mathbf{k^\prime})}{2}}\right)\Delta\left(\mathbf{k}^{\prime}\right)}{2E(\mathbf{k^\prime})}, \\
\label{Eq: Gap}
\end{equation}
where $V(\mathbf{k, k^\prime})$ is the pairing interaction, approximated by a separable potential
with an energy cutoff given by the phonon energy $\omega_0$ (in units of $\hbar = 1$) and given by $V(\mathbf{k, k^\prime}) = -V_0\Theta(\omega_0 - |\xi(\mathbf{k})|)\Theta(\omega_0 - |\xi(\mathbf{k^\prime})|)$, with $V_0$ a positive strength of interaction. Here $E(\mathbf{k}) = \sqrt{\xi(\mathbf{k})^2 + \Delta(\mathbf{k})^2}$ is written in terms of a two-dimensional tight-biding electronic dispersion $\xi(\mathbf{k}) = \epsilon(\mathbf{k}) - \mu$,

\begin{equation}
    \xi(\mathbf{k}) = -2t\left(\cos k_x + \cos k_y\right) - \mu, 
\label{Eq: dispersion}
\end{equation}
with nearest neighbor hoping $t$ and chemical potential $\mu$. The lattice spacing is set to unity, $a = 1$. We choose a dispersion without next-nearest neighbor hopping, $t^{\prime} = 0$, even though nesting and possible charge density waves effects may be present \cite{Dee2023, Perali1998}. As we shall see, we will focus in a regime of parameters that justifies this choice.  Moreover, the temperature is set by $\beta = 1/T$ (in units of $k_B = 1$) and $\Omega$ is the area of the two-dimensional $k-$space. The gap carries the same cutoff of the separable interaction $\Delta(\mathbf{k}) = \Delta_0 \Theta(\omega - |\xi(\mathbf{k})|)$. The finite-temperature density equation is written as

\begin{equation}
n=\frac{2}{\Omega} \sum_k\left[v(\mathbf{k})^2 f\left(-E(\mathbf{k})\right)+u(\mathbf{k})^2 f\left(E(\mathbf{k})\right)\right],
\label{Eq: Density}
\end{equation}
where $v(\mathbf{k})$ and $u(\mathbf{k})$ are the BCS weight functions, with $u(\mathbf{k})^2 = 1 - v^2(\mathbf{k})$ and $v(\mathbf{k})^2 = (1/2)(1-\xi(\mathbf{k})/E(\mathbf{k}))$, and $f(E(\mathbf{k}))$ is the Fermi-Dirac distribution function. For the dispersion in Eq. \eqref{Eq: dispersion}, we can connect the interaction strength $V_0$ with the dimensionless electron-phonon coupling $\lambda$ as $\lambda = NV_0$, where $N = 1/4\pi a^2t$ is the density of states at the bottom of the band. We have solved self-consistently Eqs. \eqref{Eq: Gap} and \eqref{Eq: Density} in order to obtain the phase boundaries for the BCS-BEC crossover in terms of the finite-temperature condensate fraction, $\alpha$, which is the ratio between the number of fermions forming Cooper pairs and the total number of fermions in the band, defined as \cite{Salasnich2005}

\begin{equation}\alpha=\frac{\sum_{\mathbf{k}} u(\mathbf{k})^2 v(\mathbf{k})^2 \tanh ^2 \left(\frac{\beta E(\mathbf{k})}{2}\right)}{n}.
    \label{Eq: cond fraction}
\end{equation}

This quantity evolves smoothly from $\alpha = 0$ in the extreme BCS limit to $\alpha = 1$ in the extreme BEC limit.
Following Ref. \cite{Guidini2014}, we identify the region $0.0< \alpha < 0.2$ as the  weak-coupling BCS regime, $0.2 < \alpha < 0.8$ region as the crossover, and $\alpha > 0.8$ as the strong-coupling BEC regime. 
Once the boundaries are determined for a given set of electron-phonon coupling $\lambda$, and cutoff phonon energy $\omega_0$, we are also able to determine the renormalization of the chemical potential, $\mu$, which from the BCS to the BEC phases approaches half of the binding energy of the bound state. 
This characterization of the crossover allows us to identify if the nonadiabatic corrections affect more or less strongly different regions of the crossover.

In addition to the chemical potential, another key parameter that allows us to connect the nonadiabatic correction to the BCS-BEC crossover is the temperature. Since we are dealing with a two-dimensional system, we define two important temperature scales. Indeed in superconducting thin films, by lowering the temperature the superconducting transition occurs in two stages: at first, the superconducting order parameter acquire a non-zero amplitude, but without global phase coherence, at $T_{MF}$, that is determined by the condition $\Delta(T = T_{MF}) = 0$, given by Eq. (\ref{Eq: Gap}). This temperature regime is called the pseudogap phase. This phase is characterized by the proliferation of free quantized vortices causing an exponential decay of coherence. At lower temperature, the true superconducting transition occurs at $T_{BKT}$ which is defined by the Nelson-Kosterlitz (NK) condition \cite{Koster}
\begin{equation}
T_{BKT}= \frac{\pi}{2} J (T_{BKT})
\label{eq: BKT}
\end{equation}
where $J(T)$ is the superconducting phase stiffness. It measures the energy cost associated with space variations in the phase of the order parameter. In the BCS approach, this quantity goes to zero monotonically at $T_{MF}$. However, in order to get the experimentally observed behavior of $J(T)$, which is a monotonically decreasing function of temperature, exhibiting a discontinuous jump from a finite value to zero at $T_{BKT}$ \cite{Nels}, we calculate $J(T)$ by solving the renormalization group flow equations \cite{nagaosa1999, Koster, Nels} derived by Nelson and Kosterlitz, using the BCS stiffness as an initial condition, in such a way to characterize the superconducting transition defined by $T_{BKT}$. In the temperature regime below $T_{BKT}$, the system is characterized by bound vortex-antivortex pairs and an algebraic quasi long-range order. In the next section, we set the temperature scale in such a way to guarantee that the system is located in the pseudogap phase, with temperatures in between the mean-field and the BKT transition temperatures, which allows us to span the BCS-BEC crossover in the proper regime of parameters.

\section{Numerical results}
\label{Section: numerical analysis}

\subsection{Establishing the temperature scale}
\label{subsec: A}

\begin{figure}
    \centering\includegraphics[width = \linewidth]{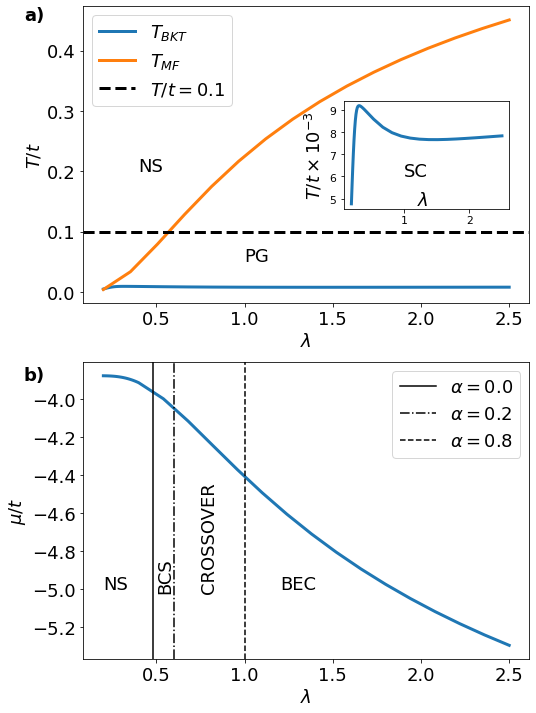}
    \caption{a) Temperature phase diagram obtained from the mean-field coupled BCS equations and the NK criterion as a function of the electron-phonon coupling $\lambda$, defining the normal state (NS), pseudogap (PG) and superconducting (SC) states. The inset shows the SC phase defined by the BKT transition temperature. The phonon frequency is set to $\omega_0/t = 1.5$ and the density is $na^2 = 0.02$. b) The chemical potential $\mu/t$ as a function of $\lambda$, together with the condensate fraction lines defined by $\alpha = 0$ for the transition from NS to BCS, $\alpha = 0.2$ from BCS to the crossover and $\alpha = 0.8$ entering the BEC regime. The phonon frequency and density are the same as in a), and the temperature is $T/t = 0.1$.}
    \label{fig:T diagram}
\end{figure}




Figure \ref{fig:T diagram}a)  shows the temperature phase diagram in terms of the electron-phonon coupling $\lambda$ obtained from the mean-field coupled equations and the NK criterion, for $\omega_0/t = 0.1$ and fixed density, $na^2 = 0.02$. We can identify the normal state (NS), pseudogap (PG) and superconducting (SC) phases of the system. Clearly, there is a large region of temperatures in the phase diagram in which the system is in the PG phase, in such a way that the coherent superconducting state is suppressed to lower temperatures, as we show in the inset. In Fig. \ref{fig:T diagram}b) we show the renormalization of the chemical potential as a function of the coupling $\lambda$, together with the associated lines of the condensate fractions, defining the boundaries of the NS, BCS, crossover and BEC regimes, for $T/t = 0.1$, as defined in \ref{fig:T diagram}a) with the black dashed line. It is clear that we can track the entire crossover, starting from the normal state, in terms of the electron-phonon coupling, staying in the pseudogap phase, where the fluctuations are driving the strong renormalization of the chemical potential below the bottom of the band even away from the superconducting state.

By analyzing the temperature phase diagram, we set the temperature scale as $T/t = 0.1$ throughout the rest of the numerical analysis. We verified that even for different frequencies and densities this temperature always places the system in the PG phase, above the BKT transition temperature, and we can trace the entire crossover by varying the electron-phonon coupling, as we can see from the renormalization of the chemical potential.

\subsection{Nonadiabatic vertex corrections}
\label{subsec: B}


Here we will deepen the analysis of the behavior of diagrams in Eqs. \eqref{Eq: P} and \eqref{Eq: C} in terms of the exchanged momenta, chemical potential and phonon frequency. When note explicitly stated, all the energy scales are measured in units of the hopping parameter $t$. For simplicity, we start by choosing a momentum-independent electron phonon coupling $g(\mathbf{q}) = g_0$, which is connected with $\lambda$ for a square lattice tight-biding dispersion as $g_0 = \pi \lambda$, and setting the external Matsubara frequencies as $\omega_k = \pi T$ and $\omega_{k^{\prime}} = \omega_0$ in order to ensure we are contemplating a dynamical regime. 

\begin{figure*}
    \centering\includegraphics[width=\textwidth]{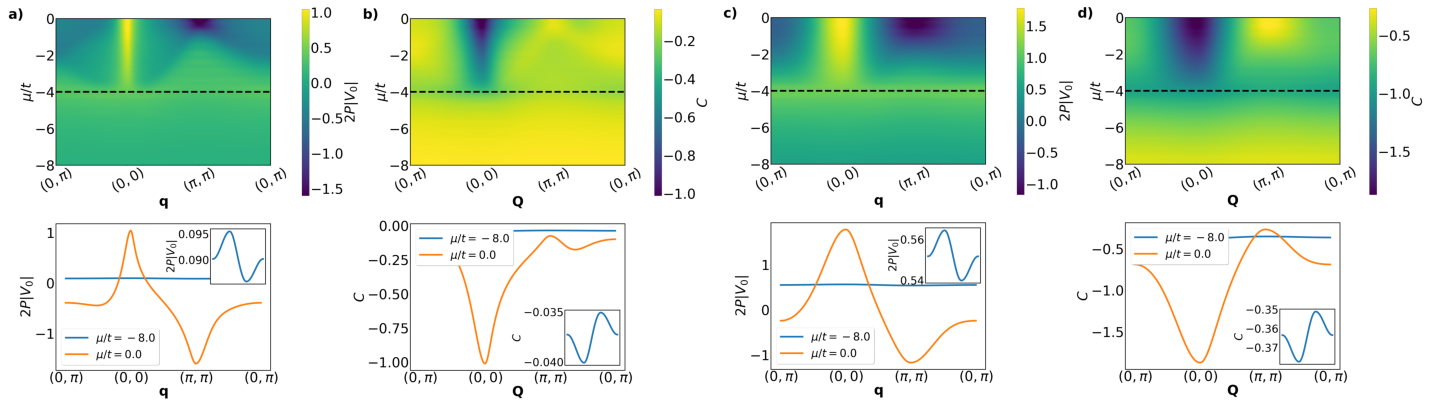}
    \caption{Vertex and cross corrections, $P$ and $C$, as a function of chemical potential, $\mu/t$, and exchanged momenta $\mathbf{q = k - k^\prime}$ and $\mathbf{Q = k + k^\prime }$ along the path $(\pi, 0) \rightarrow (0,0) \rightarrow (\pi, \pi) \rightarrow (0, \pi)$ in the first Brillouin zone. (Top) a) and b) Color maps of $2P|V_0|$ and $C$, respectively, for the intermediate coupling regime with parameters $\omega_0/t = 1.0$, $\lambda = 1.0$ and $T/t = 0.1$. (Top) c) and d) Color maps of $2P|V_0|$ and $C$, respectively, for the strong coupling regime with parameters $\omega_0/t = 4.0$, $\lambda = 2.0$ and $T/t = 0.1$. The black dashed line represents the bottom of the band ($\mu/t=-4$), while the van Hove singularity at half-filling is located at $\mu/t=0$. Bottom of all panels shows $P$ and $C$ for two selected values of $\mu/t$ with the respective parameters related to the top panel, with insets showing the limiting case $\mu/t = -8.0$.}
    \label{fig:P and C}
\end{figure*}

In Fig. \ref{fig:P and C}, we show the behavior of the vertex and cross diagrams across a chosen path in the first Brillouin zone, characterizing the exchanged momenta $\mathbf{q}$ and $\mathbf{Q}$, for different values of the chemical potential $\mu/t$, going from the half-filling case, $\mu/t = 0.0$, where the van Hove singularity is located, towards negative values below the bottom of the band $\mu/t < -4.0$. For Figs. \ref{fig:P and C}a) and b) we choose the intermediate coupling regime, with parameters $\lambda = 1.0$ and $\omega_0/t = 1.0$, and for c) and d) the strong coupling regime with $\lambda = 2.0$ and $\omega_0/t = 4.0$. For the case of $P$, we report the quantity $2P|V_0(k-k^\prime)|$ in order to be consistent when comparing the contributions from vertex and cross corrections. This choice can be understood by looking at Eq. \eqref{Eq: effective}, where the overall contribution of $P$ is multiplied by a negative $2V_0(k-k^\prime)$ factor to acquire an energy unit, while $C$ itself already has units of energy. In Fig. \ref{fig:P and C} we see that when the chemical potential is inside the band, both the corrections acquire a strong dependence on momenta, specially for small scattering vectors around $\mathbf{q} = (0,0)$, where $P$ displays a maximum and $C$ a minimum, which in the context of the dressed effective interaction in Eq. \eqref{Eq: effective}, enhances the pairing. 
This is in agreement with the picture of a momentum dependent electron-phonon coupling peaked at small momenta, which selects mainly forward scattering, that is usually found in small carrier density systems, as the one here considered, and in other strongly correlated systems. 
Furthermore, around $\mathbf{q} = (\pi, \pi)$ and $\mathbf{Q} = (\pi, \pi)$, we see a strong dependence on band filling for both corrections, which is associated with nesting effects and the appearance of strong CDW correlations \cite{Dee2023}. This effect can be suppressed by the inclusion of a next nearest neighbor hopping $t^\prime$ \cite{Perali1998}. Yet, our focus lies on features near $\mathbf{q} = (0,0)$.

Interestingly, the evolution towards the strong coupling regime broadens the peaks around $(0,0)$ and $(\pi, \pi)$, which enables more scattering vectors to contribute to the overall corrections, which also displays larger values since both are proportional to $\lambda$. The lower panel of Fig. \ref{fig:P and C} shows the behavior of the corrections for the two limiting cases we explore, $\mu/t = 0.0$ at half-filling and $\mu/t = -8.0$, deep outside the band. We see that the strong dependence of the corrections in momenta are suppressed by the decrease in the chemical potential, even though the overall trend remains, which is peaked at small momentum transfer around $\mathbf{q} = (0,0)$ and a combination of positive $P$ and negative $C$. As we shall see below, the combination of these two features is important for the analysis of the corrections in the BCS-BEC crossover.

We now turn our attention to the overall effect of the corrections in the interaction potential, as in Eq. \eqref{Eq: effective}. Since $P$ is a function of $\mathbf{q = k - k^\prime}$ and $C$ of $\mathbf{Q = k + k^\prime}$, the dressed pairing interaction $V(\mathbf{q, Q})$ has a complex behavior as a function of these exchanged momenta. To overcome this difficulty, we choose two distinct cases: forward scattering, $\mathbf{k = k^\prime}$, which sets $\mathbf{q = 0}$ and $\mathbf{Q} = 2\mathbf{k}$ and back-scattering, $\mathbf{k = -k^\prime}$, which gives $\mathbf{Q = 0}$ and $\mathbf{q} = 2\mathbf{k}$. In both cases, the non-zero value of the exchanged momenta is determined by the Fermi vector defining the Fermi surface in the $\Gamma - Y$ direction, whenever the chemical potential is inside the band. For example, for the back-scattering case, $\mathbf{q} = (0, 2k_F^y)$, where $k_F^y$ is the component of the Fermi vector along the $y$ axis, defined for each $\mu/t$ using the dispersion in Eq. \eqref{Eq: dispersion}. When the chemical potential is below the bottom of the band, since no Fermi surface is defined, we set the exchanged momenta to zero in both cases.

In Fig. \ref{fig:mu-omega diagram} we show the dimensionless effective interaction relative to the bare interaction $V_0$, which we define as 
\begin{equation}
V_{eff}(\mathbf{q, Q} ) = \frac{V(\mathbf{q, Q} )- V_0}{V_0},
\end{equation}
to visualize amplification effects whenever $V_{eff} > 0$, as a function of phonon frequency $\omega_0/t$, for different chemical potentials. For $\mu/t \leq -4.0$ there is no difference between forward and back-scattering processes, but for chemical potentials inside the band, we show the representative case of $\mu/t = -2.0$ to differentiate between the two. For the forward scattering case, we get an amplification of the order of $80\%$ inside the band. Remarkably, as a function of the phonon frequency, the effective pairing interaction display a non-monotonic behavior, with a maximum exhibiting a large amplification of nonadiabatic effects for relatively small values of $\omega_0/t$. For the back-scattering case, however, the trend is different. The amplification  inside the band is qualitatively comparable with the one caused when the chemical potential is below the bottom of the band. Interestingly, the largest enhancement of pairing occurring for $\mathbf{k = -k^\prime}$ is exactly at the bottom of the band, $\mu/t = -4.0$. This can be traced back to the non-trivial behavior of the vertex function $P$ as a function of exchanged momenta $\mathbf{q}$. As the chemical potential is pushed outside the band, the Fermi surface is smeared, the exchanged momentum is continuously decreased, favoring enhancements of the effective pairing due to the nonadiabatic corrections. In this scenario, the contribution arising from $P$ are more relevant than the ones arising from $C$. It is also interesting to note that as the system approaches the BEC with $\mu/t < -4.0$, in order to have a stabilization of the nonadiabatic effects, one needs a large phonon frequency, since for small $\omega_0$, the pairing window does not encompass the band and the effects of corrections are small.

\subsection{Mean-field BCS-BEC crossover in the nonadiabatic regime}

\begin{figure}[!t]
    \centering\includegraphics[width = \linewidth]{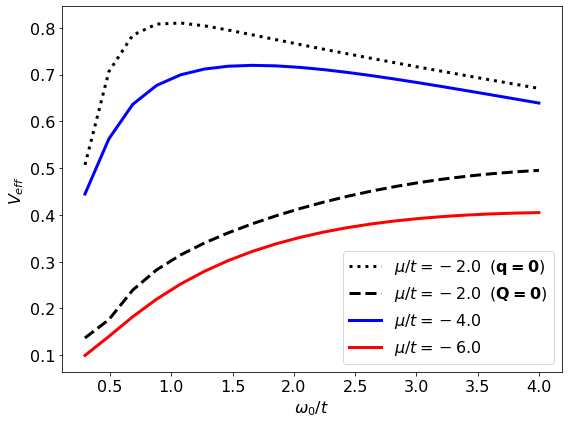}
    \caption{ Effective interaction as a function of phonon frequency $\omega_0/t$, relative to the bare interaction, for selected values of the chemical potential $\mu/t$ for forward scattering ($\mathbf{q=0}$), and back-scattering ($\mathbf{Q=0}$). The black lines are for $\mu/t$ inside the band, red for outside, and blue at the band bottom. The other parameters are set as $\lambda = 2.0$ and $T/t = 0.1$.}
    \label{fig:mu-omega diagram}
\end{figure}

\begin{figure}[!h]
\centering\includegraphics[width = \linewidth]{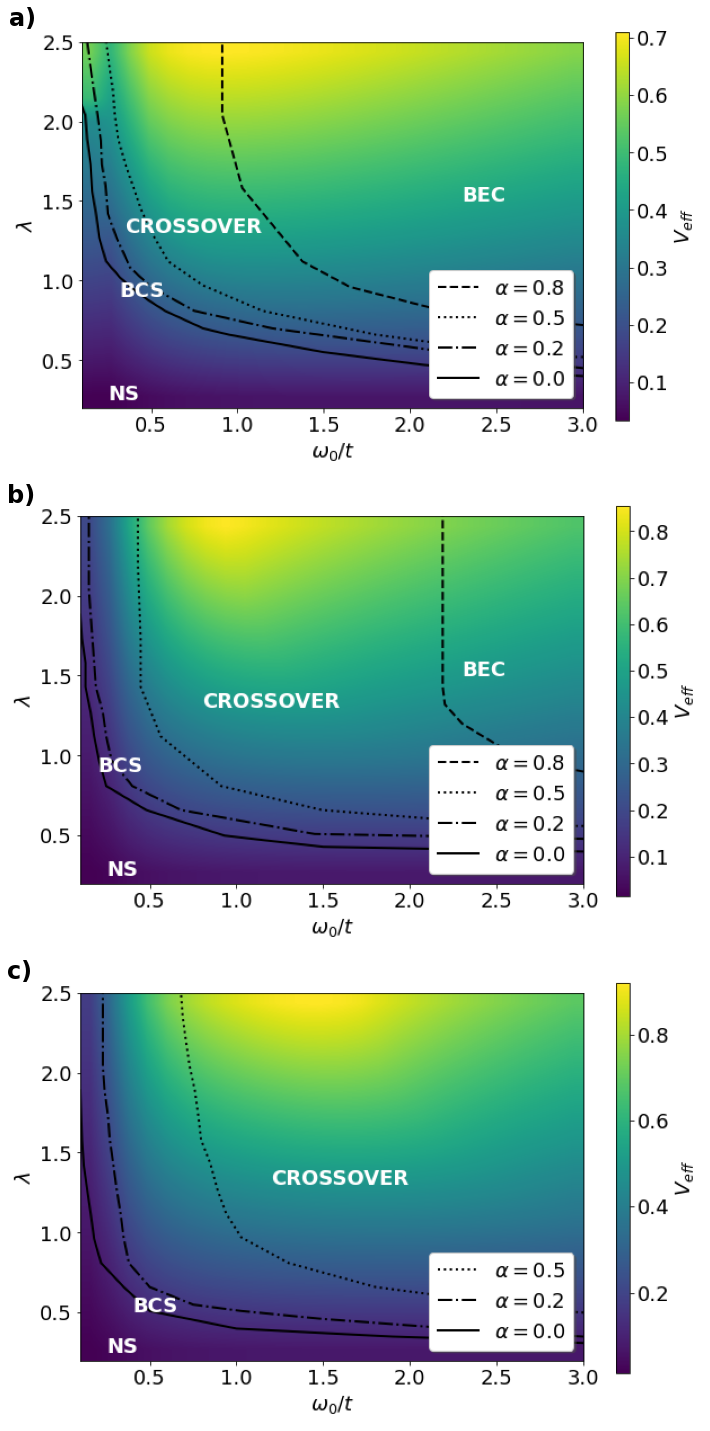}
    \caption{Nonadiabatic effective interaction in the $\omega_0/t - \lambda$ parameter space obtained from the mean-field coupled equations, together with the NK criterion, for three values of density: a) $na^2 = 0.02$, b) $na^2 = 0.05$  and c) $na^2 = 0.1$. The curves represent the condensate fraction, $\alpha$, defining the boundaries for the normal state (NS) ($\alpha = 0$), BCS ($0<\alpha < 0.2$), crossover ($0.2 < \alpha < 0.8$) and BEC ($\alpha > 0.8$). We also display the $\alpha = 0.5$ right in the middle of the crossover. The other relevant parameters were set to $\omega_k = \pi T$, $\omega_{k^\prime} = \omega_0$ and fixed $T/t = 0.1$ for all $(\omega_0/t, \lambda)$ pairs.}
    \label{fig:Veff_diagram}
\end{figure}

Here we establish a connection between the mean-field BCS-BEC equations discussed in Sec. \ref{subsec: A} and the nonadiabatic effects in Sec. \ref{subsec: B}, which is summarized in Fig. \ref{fig:Veff_diagram}, where we show the effective interaction $V_{eff}$ in the $\omega_0/t-\lambda$ parameter space, together with the curves defined by $\alpha = 0.0$, $\alpha = 0.2$, $\alpha = 0.5$ and $\alpha = 0.8$, for different densities, namely $na^2 = 0.02$ in Fig. \ref{fig:Veff_diagram}a), $na^2 = 0.05$ in b) and $na^2 = 0.1$ in c). Moreover, we use the values of $\mu/t$ coming from the self-consistent mean-field equations for the calculation of the vertex corrections. Motivated by the results in Figs. \ref{fig:T diagram}, \ref{fig:P and C} and \ref{fig:mu-omega diagram}, we choose the forward scattering sector $\mathbf{q} = 0$ and $\mathbf{Q} = 2\mathbf{k_F}$ for the calculation of the effective interaction, following the same prescription as above discussed for setting the exchanged momenta, since it is in this parameter regime where we have the greatest enhancements of nonadiabatic effects.

In terms of the BCS-BEC crossover in a single-band system, the carrier density plays a crucial role. By using the condensate fraction as a marker to locate the system throughout the crossover, the results reported in Fig. \ref{fig:Veff_diagram} show that, for the parameters analyzed here, the BEC is slowly suppressed as the density of the system is increased. Overall, and with the parameters relevant for each side of the crossover, the effects of nonadiabaticity are to increase the effective interaction. Remarkably, one the most important results of the present analysis is that for all the densities considered, the largest amplifications start to occur when half of the particles in the system are forming the condensate, in the middle of the crossover. This happens for relatively small values of phonon frequency $\omega_0 \sim t$ and large electron-phonon coupling $\lambda$. This can be traced back to the behavior of the nonadiabatic diagrams as the chemical potential is pushed outside the band bottom and the selected exchanged momenta are small, as it happens in the crossover and discussed below. 

From the results in Fig. \ref{fig:Veff_diagram}, it is clear that nonadiabatic effects are relevant and must be taken into account, for instance in quasi-flat band systems in the presence of strong electron-phonon coupling.
As an example, we consider graphene moiré superlattices in the form of  magic-angle-twisted bilayer and trilayer graphene (MATBG and MATTG, respectively). 
In these systems, electron-electron correlations, electron-phonon coupling and nonadiabatic effects are expected to be enhanced by the existence of small Fermi energy flat bands \cite{Cao2018}. An important role of electron-phonon coupling has been indeed proposed early on for MATBG \cite{Angeli2019}, but the nonadiabatic effects on superconductivity have not been carefully explored, to the best of our knowledge.
From Refs. \cite{Choi2021, Koshino2018}, we can extract the electron-phonon coupling $\lambda$ for specific phonon modes at the $\Gamma$ point of the Brillouin zone, with characteristic frequency $\omega_0$ and the effective hopping parameter, $t$, connecting the spots in the moiré pattern (which emerges as a honeycomb lattice). Due to the large ratio $\omega_0/t$ because of the flatness of the bands, these systems are located out of the parameter regime explored in the phase diagrams in Fig. \ref{fig:Veff_diagram}, however we can argue that the effects of nonadiabaticity induce relevant amplifications up to $30\%$ in the pairing. Moreover, we point out that in these systems, depending on the value of the Fermi energy, the electron-phonon coupling can be as large as $\lambda = 2.0$, which would place this family of materials in a region of even larger amplification of the pairing due to nonadiabatic effects \cite{Choi2021}. Interestingly, the superconductivity in MATTG is most probably of strong-coupling nature and can be electrically tuned to be close to the BEC region of the two-dimensional BCS-BEC crossover, as experimentally observed \cite{Park2021} and captured by our calculations.

We extend the connection with real systems by analyzing superconducting fullerides. From Ref. \cite{Han2003} we can extract typical parameters for alkali-doped $A_{x}C_{60} \; (A = K, Rb, Cs)$, namely $\lambda \sim 0.5 - 1.0$ and $\omega_0/t \sim 0.8 - 2$ to locate the system in our phase space for pairing amplification. If we choose $\omega_0/t = 1.2$ and $\lambda = 0.8$, we see that it lies within the crossover and above the $\alpha = 0.5$ line, indicating that the strong coupling nature of the superconductivity drives the system towards the BEC side of the crossover, where the nonadiabatic effects are most important. Fullerides are known to be strongly correlated superconductors \cite{Capone2002,Ganin2008,Capone2009}, with a robust Mott insulator phase, thus a precise analysis should take into account not only strong electron-electron interactions, but also the possible momentum dependence of the electron-phonon coupling matrix, due to the presence of several phonon modes \cite{Nomura2016}. However, since intramolecular lattice vibrations are the main contributors to the total electron–phonon coupling \cite{Schluter1992} and they show little dispersion in momentum \cite{Nomura2015}, an estimation of nonadiabatic corrections can safely neglect these dependencies.

\section{Conclusions}
\label{Section: Conclusion}

In this work, we have undertaken an extensive analysis of the vertex and cross correction diagrams in the superconducting self-energy, particularly in the context of the nonadiabatic regime, as the system is driven towards the BCS-BEC crossover. By employing mean-field BCS equations to solve for the superconducting pseudogap and density, together with the NK criterion for determining  the BKT transition temperature, we construct the temperature phase diagram in terms of the electron-phonon coupling. We locate our model system inside the PG phase and study the renormalization of the chemical potential as the system goes from the unpaired normal state towards the BCS, crossover and BEC regimes. We show that the BKT transition  temperature is orders of magnitude smaller than the PG onset temperature and, therefore, the pairing fluctuations in the normal state cause the renormalization of the chemical potential. We then determine the dependence of the corrections on exchanged momenta and verify that it diminishes with the renormalization of the chemical potential directed below the bottom of the electronic band, which is a central characteristic of the BCS-BEC crossover. Despite this, we identify persistent amplifications to the effective pairing. We further scrutinize the limits of forward and back-scattering of pairing electrons, revealing that strong enhancements occur in the forward-scattering sector for relatively small phonon frequencies and large electron-phonon coupling. Intriguingly, in the back-scattering scenario, the behavior of the effective interaction remains qualitatively unaltered both inside and outside the band. On top of that, we connect the mean-field BCS-BEC physics with the nonadiabatic corrections by mapping the relevant parameters to the calculation of the effective pairing interaction in the presence of the vertex and cross diagrams, unveiling large amplifications across the entire BCS-BEC crossover. Notably, we observe that these enhancements are more pronounced when over half of the particles are in the incoherent condensate state, corresponding to the middle of the crossover. 

For strong electron-phonon coupling and for the chemical potential approaching the bottom of the band, the perturbation theory on the parameter $\lambda \omega_0/E_F$, may become unreliable, since the system can reach (bi)polaronic instabilities \cite{Capone1998,Alexandrov2000}. However, the perturbative approach is a good starting point to understand the behavior of the nonadiabatic corrections in the BCS-BEC crossover from the normal metallic state, where the electron-phonon coupling opens new interacting channels, but it retains its metallic behavior. Moreover, it is not clear if the small parameter remains the same when the chemical potential goes below the bottom of the band. Our numerical results suggest that there is no divergent behavior of $V_{eff}$, which remains always less than unity, not even doubling the bare interaction $V_0$. This shows that the nonadiabatic diagrams are not diverging. Therefore, we can get qualitative understanding of the nonadiabatic corrections in the BCS-BEC crossover even beyond the standard perturbative range of validity \cite{Deppeler2002, Capone1998, Botti2002}. 

Even in the simplified approach used here, we locate some specific systems, namely MATBG, MATTG and alkali-doped fullerides in the phase diagrams. For the parameters characterizing these systems, we show that the presence of nonadiabatic pairing effects induce large pairing amplifications, which can lead to enhancements of the superconducting critical temperature. Moreover, the carrier density can be fine-tuned in the aforementioned systems, so their precise location in the crossover in terms of the condensate fraction of incoherent pairs or of the chemical potential may vary, even though drastic changes in $\lambda$ and $\omega_0$ are not expected. Furthermore, due to their quasi-flat band nature, presence of van Hove singularities and large electron-phonon coupling, we argue that nonadiabatic corrections can also be important in the study of superconductivity in kagome metals \cite{Wu2021, Zhong2023}. We point to future investigations that can generalize the vertex corrections to multiband systems, to understand the effects of screening of pairing fluctuations \cite{Pisani2018, Salasnich2019, Tajima2019} and the inclusion of the anomalous diagrams in the superconducting self-energy \cite{Botti2002} to study the BCS-BEC crossover. In multi-orbital systems, the electron-phonon coupling also competes with the Hund's exhange coupling\cite{Capone2002}, leading to effects that can enhance the role of the nonadiabatic effects \cite{Scazzola2023}. 

\section*{Acknowledgments} 
We acknowledge Marcello B. Silva Neto for useful comments. This work has been supported by PNRR MUR project PE0000023-NQSTI. M.C. also acknowledges  financial support of MUR via PRIN 2020 (Prot. 2020JLZ52N 002) programs, PRIN 2022 (Prot. 20228YCYY7), National Recovery and Resilience Plan (NRRP) MUR Project  No. CN00000013-ICSC. The views and opinions expressed are solely those of the authors and do not necessarily reflect those of the European Union, nor can the European Union be held responsible for them.

\newpage

%

\end{document}